\documentclass[twocolumn]{revtex4}%
\usepackage[paperwidth=210mm,paperheight=297mm,centering,hmargin=2cm,vmargin=2.5cm]{geometry}
\usepackage{amsfonts}
\usepackage{amsmath}
\usepackage{amssymb}
\usepackage{bm}
\usepackage{graphicx}
\usepackage{color}
\usepackage{soul}
\usepackage{epsfig}%
\usepackage[dvipsnames]{xcolor} 
  \definecolor{bleu_cite}{RGB}{0,0,255}
\usepackage[colorlinks=true,allcolors = black,citecolor=bleu_cite,  ]{hyperref} 
\usepackage{natbib}

\begin{document}
\title{Dynamical control over the confinement of spatially indirect excitons in electrostatic traps of GaAs coupled quantum wells}
\author{Mussie Beian$^{1,3}$, Suzanne Dang$^{1}$,  Mathieu Alloing$^{1,3}$, Romain Anankine$^{1}$, Edmond Cambril$^2$, Carmen Gomez$^2$, Johann Osmond$^3$, Aristide Lema\^{i}tre$^2$ and Fran\c{c}ois Dubin$^{1}$} 
\affiliation{$^1$ Institut des Nanosciences de Paris, CNRS and UPMC, 4 pl. Jussieu,
75005 Paris, France}
\affiliation{$^2$ Laboratoire de Photonique et Nanostructures, LPN/CNRS, Route de
Nozay, 91460 Marcoussis, France}
\affiliation{$^3$  ICFO-Institut de Ciencies Fotonicas, The Barcelona Institute of Science and Technology, 06880 Castelldefels, Spain}

\date{\today }

\begin{abstract}
We study spatially indirect excitons confined in a 10 $\mu$m wide electrostatic trap of a GaAs double quantum well.  We introduce a technique to control the amplitude of the electric field interacting with the excitons electric dipole,  with nanosecond precision. Our approach relies on electronic waveforms corrected for the distorsions occurring at highest frequencies so that impedance matching is not necessary. Thus, we manipulate the confinement of cold gases without inducing sizeable perturbations down to sub-Kelvin bath temperatures.  
\end{abstract}

\pacs{XX}
\maketitle

\section{Introduction}

Spatially indirect excitons are long-lived semiconductor quasi-particles which constitute a model system in order to explore collective quantum phenomena \cite{Monique_review}. They are engineered by enforcing a spatial separation between their electrons and holes constituents, a geometry which is conveniently achieved using two coupled quantum wells subject to a perpendicular electric field \cite{sivalertporn2012direct}. Thus, minimum energy states for electrons and holes lie each in a distinct layer, indirect excitons resulting from the Coulomb attraction between opposite-charge carriers. By imposing such spatial separation one provides indirect excitons with a large permanent electric dipole and unique physical properties to probe cold gases. For example, electrons and holes confined in coupled quantum wells have wave functions with a reduced overlap, leading to optical lifetimes typically around 100 ns for indirect excitons. Combined to a nanosecond cooling in two-dimensional heterostructures \cite{ivanov2004thermalization,Oh_2000}, this lifetime allows studying gases at thermal equilibrium with the semiconductor matrix, down to sub-Kelvin temperatures \cite{Beian_2015}.

To possibly explore the exotic many-body phases predicted below a few Kelvin for indirect excitons  \cite{Lozovik_1975,Blatt_1962,Keldysh_BEC,Monique_BEC,Combescot_2012}, it is crucial to control their spatial confinement. This requirement may be achieved by controlling the strength of the field interacting with the large electric dipole characterizing indirect excitons, oriented perpendicular to the plane of the bilayer heterostructure. 
To this aim, a new technology has emerged about ten years ago. It relies on nano-fabricated gate electrodes deposited at the surface of a field-effect device embedding two coupled quantum wells (CQW) \cite{high2009trapping,gartner2007micropatterned,schinner2013confinement,chen2006artificial,gorbunov2006large}. By patterning the shape of the surface electrodes, one can manipulate the electric field applied perpendicular to a CQW and thus trap indirect excitons in the regions where the field is the strongest \cite{Beian_2015,schinner2013confinement,Kuznetsova_2015,High_Nano_Lett_2009,Timofeev_2011}, or even control the exciton transport  \cite{Winbow_2011,High_2008,Hasling_2015}.

While rapid progress has been achieved for the static control over the excitons confinement, the dynamical control of electrostatic traps has evolved less. So far, it has only been achieved by adapting the very large impedance of a field-effect device embedding a CQW, i.e. by placing it in parallel to a calibrated impedance (typically 50 $\Omega$) \cite{winbow2008photon}. Combined to electrical transmission lines, this approach ensures that the shape of short electrical pulses is efficiently transferred towards an electrostatic trap cooled in a helium cryostat. Thus, the internal electric field acting on indirect excitons is varied within nanoseconds. 

Although impedance matching is efficient from the electronic point of view, it is not very suited to low temperature measurements. Indeed, this approach leads to thermal heating since a few D.C. volts are typically necessary to control exciton traps, thus dissipating $\sim$100 mW if the impedance of a field-effect device is matched to 50 $\Omega$. Such heating can not be compensated cryogenically in the sub-Kelvin regime, i.e. in the regime where indirect excitons can realise quantum gases \cite{Anankine_2017}. Accordingly, the frontiers that can be addressed experimentally are limited. As a concrete example, let us first note that exciton cooling by acoustic phonons is inefficient at sub-Kelvin temperatures \cite{Oh_2000,ivanov2004thermalization}. This limitation poses intriguing questions regarding the thermalisation in this regime and the underlying role of exciton-exciton interactions. Interestingly, this open problem could be addressed by varying dynamically the strength of an electrostatic trap confining indirect excitons, in a fashion similar to evaporative cooling schemes routinely utilised for atomic gases \cite{Ketterle_1996}. Moreover, a dynamical control over the confinement of indirect excitons, free from additional heating, constitutes a crucial requirement to explore innovative schemes to couple exciton and polariton condensation \cite{Shelykh_2015}. Ultra-fast electric-field tuning might also find applications in quantum information science to control dynamically the emission/absorption energy of neutral or charged quantum dots \cite{Vuckovic_2012}.

Here we report experiments highlighting that electrostatic traps  have a confinement strength that can be varied with nanosecond precision, without relying on impedance matching. We show that arbitrary temporal waveforms, e.g. square pulses, can be programmed in-situ once the transfer function of the electronic circuit, including all passive elements, has been quantified.  Limiting factors such as signal overshoot or capacitive-like slow-down are then overcome by reverse engineering. In this way, we vary the strength of the exciton confinement by about 2 meV at 330 mK, i.e. about 10 times the excitons homogeneous broadening \cite{Anankine_2016}, with nanosecond time precision and without inducing measurable perturbations such as heating. Higher variations are also accessible, up to 10 meV, but at the cost of a transient current which alters strongly the electrostatic environment explored by excitons. 

\begin{figure}\label{fig1}
\centerline{\includegraphics[width=.5\textwidth]{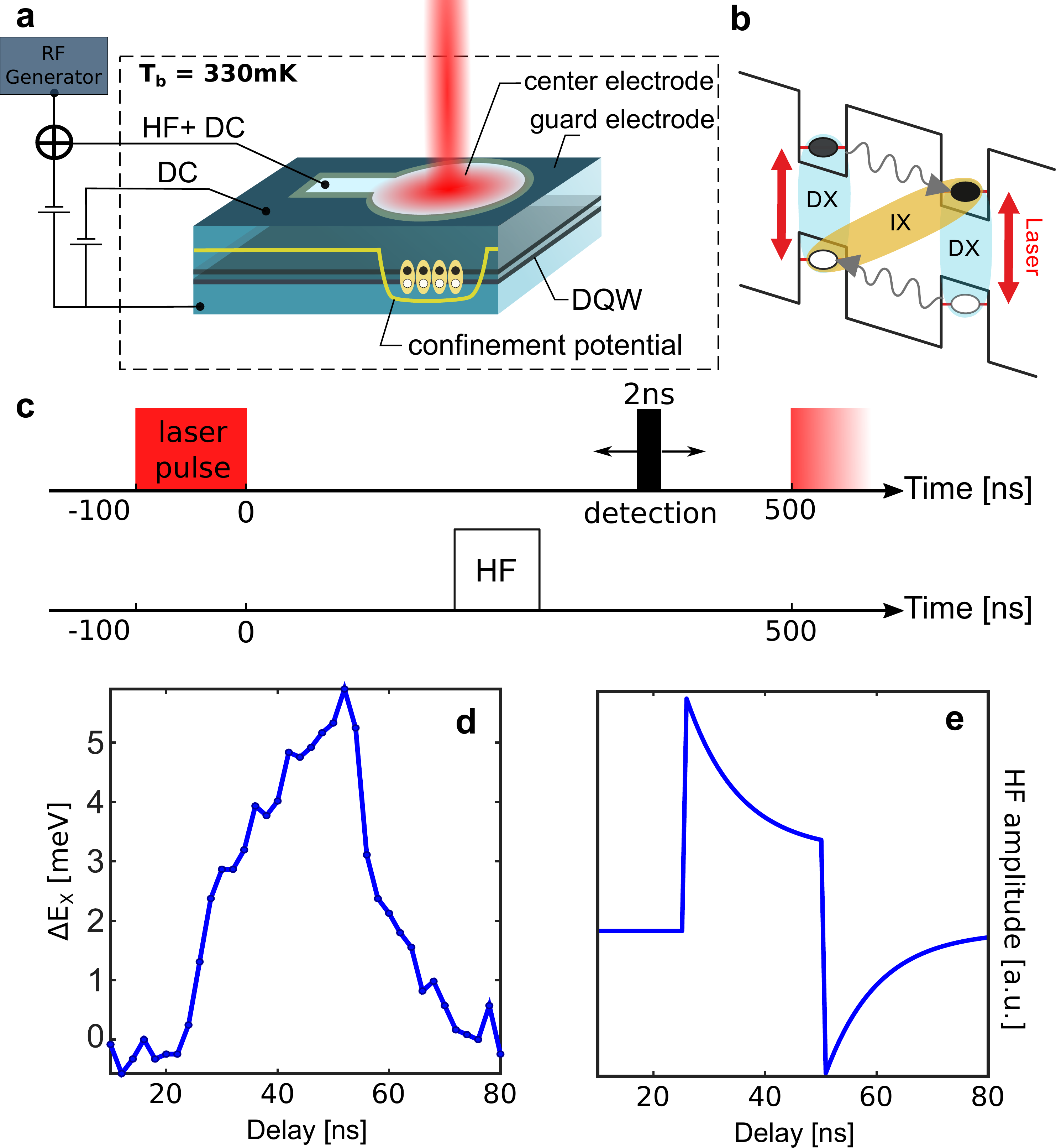}}
\caption{(a) The exciton trap consists of two independent electrodes, the centrer (trap) electrode surrounded by a guard electrode. The former gate is controlled by a DC bias mixed with a high-frequency (HF) component, provided by an analog waveform programmer. Electronic carriers are injected optically in the trap, by a laser beam covering the entire area. (b) Indirect excitons (IX) result from the tunnelling of  electrons and holes after their optical injection by a laser pulse resonant with the direct exciton absorption (DX). (c) Our experimental sequence is repeated at 2 MHz and incorporates optical injection and detection of indirect excitons (top), together with HF electrical pulses applied onto the central trap gate (bottom). (d) Variation of the excitons energy $\Delta E_X$ induced by a 30 ns long and square pulse with 0.5 V amplitude. The pulse is applied 30 ns after extinction of the laser pulse loading indirect excitons in the trap at a bath temperature T$_\mathrm{b}$=330 mK. (e) Input waveform so that $\Delta E_X$ follows an ideal square pulse of 30 ns duration.}
\end{figure}

\section{Experimental details}

As illustrated in Fig.1.a, we study a field-effect device embedding two 8nm wide GaAs quantum wells, separated by a 4 nm AlGaAs barrier. The heterostructure is connected to a custom designed printed circuit board mounted onto the He$^3$ insert of a closed-cycle He$^4$ cryostat. Our device is thus cooled to a bath temperature T$_\mathrm{b}$=330 mK. Moreover, Fig.1.a shows that we study an exciton trap relying on two surface gate electrodes, namely a central 10 $\mu$m wide disk which is separated by 400 nm from an outer wide electrode. In the following we refer to the central electrode as the trap gate and the second electrode as the guard gate. In our experiments, these electrodes are polarised independently at $V_t$ and $V_g$. These potentials consist of a static part, -4.8 V and -3.5 V for the trap and guard electrodes respectively, which is mixed with a high-frequency (HF) component for the trap electrode only, with an amplitude of at most 1.5V.

In Fig.1.c we detail the  sequence used in the following experiments. It incorporates laser and electrical pulses. Precisely, we use a 100 ns long laser pulse to load the electrostatic trap with indirect excitons, in a trapping regime ($|V_t|>|V_g|$). The laser pulse is obtained by electro-optic modulation of a CW laser diode tuned on resonance with the direct exciton absorption of the two quantum wells (at about 794 nm see Fig.1.b). The laser light has a fixed power equal to 2 $\mu$W and covers the trap electrode area. At a controlled delay to the extinction of the loading laser pulse, we add a voltage pulse to the trap gate only and monitor the electric field induced in the field-effect device, through the energy E$_\mathrm{X}$ of the photoluminescence emitted by indirect excitons. Indeed,  E$_\mathrm{X}$ reflects directly the strength of the electric field component perpendicular to the CQW  $F_z$, since $E_{X} \sim -dF_z$ where $d$ denotes the dipole moment of indirect excitons, of about 12 nm in our device. Note that the photoluminescence signal is recorded by an imaging spectrometer coupled to an intensified CCD camera with a time resolution set to 2 ns for the following experiments.

\section{Results}

Figure 1.d presents the variation of $E_X$ when a  square voltage pulse  is applied to the trap gate. We note directly that the induced time profile does not reproduce the input signal. Instead it exhibits strong distorsions with rise and fall times both of the order of a few tens of nanoseconds. Nevertheless, the variation of $E_{X}$ amounts to about 5 meV, which reveals that $F_z$ is efficiently varied by the voltage pulse. This can seem surprising at first because the input pulse is almost completely reflected at the surface of our device, due to the mismatch between the large impedance of the field-effect device and the 50 $\Omega$ coaxial cables connecting our sample to an analog waveform programmer. In fact, the variation of $E_{X}$ signals the coupling between the excitonic dipole and the electric field evanescent in the heterostructure. Indeed, the CQW is only positioned  900 nm away from the gate electrodes. This distance is small compared to the attenuation length at high frequencies thus resulting in a strong coupling, nevertheless for short pulses ($\sim$10-50 ns) $E_{X}$ is always varied $\sim30-50\%$ less than its value obtained with a static polarisation.

To optimise the dynamical control of $E_\mathrm{X}$ we computed the electrical transfer function for our experiments, from the response of the photoluminescence energy to an input square pulse. By inverting the transfer function we deduced the input signal that shall be applied onto the trap gate so that $E_{X}$ varies as a square function, with rise and fall times that are ultimately limited by the temporal resolution of our detection. The profile of such a corrected input signal is shown in Fig.1.e for the transfer function obtained from the measurements shown in Fig.1.d. We note that the corrected input signal exhibits both overshoot and down shoot, at the rising and falling edges respectively, to correct for high-frequency losses and distorsions occurring along the transmission of the electric pulse. 

\begin{figure}\label{fig2}
\centerline{\includegraphics[width=.5\textwidth]{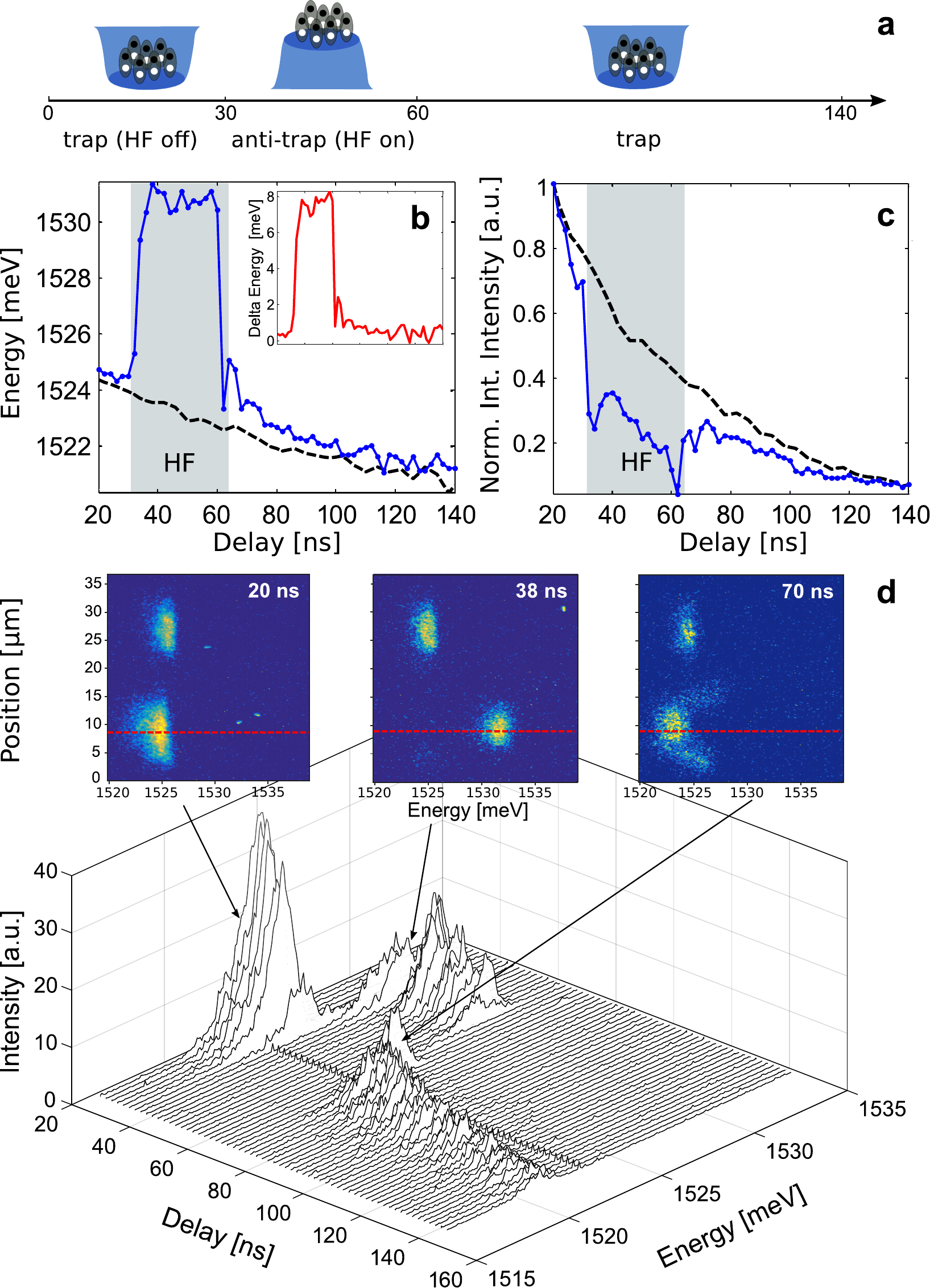}}
\caption{(a) Schematic view of the electrostatic confinement, evolving from a trap to an anti-trap while the central gate is pulsed electrically. Energy (b) and normalised integrated intensity (c) of the photoluminescence emitted from the trap; gray regions mark the timing of the square (HF) pulse applied onto the trap gate. The black dashed line shows a reference for 0V amplitude, while the amplitude is set to 1.5 V for the measurements shown by the solid blue line. In (b) the inset displays the energy difference $\Delta E_X$  between these two limit cases. (d) Spatially resolved spectra at three delays: before (20 ns), during (38 ns) and after (70 ns) the voltage pulse is applied. Experiments were realised at T$_\mathrm{b}$=330 mK with two laser spots: one focused on the center electrode (bottom spot), the other one on the guard electrode (top). Below we display photoluminescence spectra measured across the center of the trap, as shown by the dashed line in the above images, during the experimental sequence.}
\end{figure}


Figure 2 reports experiments where a corrected 30 ns long pulse is applied to the trap gate, starting from 30 ns after extinction of the loading laser. This measurement is compared to a reference situation where the electrical pulse is applied with zero amplitude. In the latter case $E_X$ decays monotonously, reflecting the decrease of the population in the trap due to radiative recombination, with a characteristic timescale of the order of 100 ns.  Remarkably, Fig. 2.b shows that the dynamics of $E_X$ is only modified during the targeted time interval when a corrected voltage pulse is applied to the trap gate. In particular, $E_X$ is varied by as much as 5 meV within 2 ns, a level of performance which is actually limited by the time resolution of our detection. 

We illustrate further the dynamical control of the excitons confinement in Fig.2.d where we report experiments including two laser excitations. As previously, a first beam excites the trap region, while a second beam is positioned a few tens of micrometers away from the trap, on the guard gate. Thus, we first verify that the electrical pulse applied onto the trap gate does not affect the exciton confinement under the guard. Also, Fig. 2.d shows that  the voltage pulse makes the confinement evolve from a trap regime, i.e. when the electric field is stronger under the trap electrode, to an anti-trap regime where the photoluminescence energy is higher under the trap gate, and finally back to a trapping configuration. Overall, the excitons energy is then varied electronically by about 8 meV during this sequence. However, as shown in Fig.2.c, the photoluminescence intensity also varies strongly when the trap gate is pulsed. Particularly, we note transient losses of photoluminescence, coincident with the two edges of the voltage pulse, at the falling edge the photoluminescence even vanishes. 

The transient losses of photoluminescence may first be attributed to a thermal heating of the heterostructure, and therefore of the trapped gas too. Indeed, the photoluminescence is only emitted by optically active states, i.e. excitons with a kinetic energy lower than about 1.5K. To asses the presence of such direct thermal heating we performed the same experiments as in Figure 2, but at a bath temperature of 2.5K. We then observed intensity losses at the pulse edges, with amplitudes similar to the ones observed at 330 mK. Accordingly, we conclude that thermal heating of the heterostructure is not dominant in our experiments. Alternatively, intensity losses may be the result of a transient current induced by the sudden variation of the internal electric field. Such a transient current would  vary the occupation of the lowest energy and optically active exciton states through collisions between cold excitons and additional free electrons/holes. In a limit where hot carriers are injected one can not exclude that the photoluminescence transiently vanishes. 

Measuring directly a transient current is unfortunatelly technically inaccessible. Indeed, Fig.2.c shows that photoluminescence losses last about 10 ns while the average photocurrent amounts to about 100 pA in these experiments. Resolving current variations in such a short time interval then requires a transimpedence of about 10$^6$ with a bandwidth greater than 100 MHz, well beyond state-of-the-art technology.  To test the presence of a transient current, we then varied the amplitude of the voltage pulse applied onto the trap gate while monitoring the variation of photoluminescence intensity at the pulse edge. We observed that the intensity drops rapidly above  0.4 V, supporting that a transient current may be activated while the internal electric field is pulsed. 

Comparing the experiments where the trap gate is electrically pulsed to the reference situation where it is not, we finally note in Fig.2.c that after its transient decrease the photoluminescence intensity is smaller in the trap during the period of electrical pulsing. This behaviour is surprising since the photoluminescence intensity is not expected to decrease when the exciton energy is varied by a few meV. However, the transient current induced in these experiments can directly increase the density of excess electrons that interact with indirect excitons in the trap. Such variation would then alter the photoluminescence yield. Indeed, recent measurements \cite{Holleitner_2012} have underlined that the photoluminescence intensity is decreased when indirect excitons interact with excess electrons. Although our experiments are not realised in a static regime as in Ref.\cite{Holleitner_2012}, the behaviour observed in Fig.2.c reproduces qualitatively the same trend and underlines further the difficulty to quantify the origin of a photoluminescence decrease at low temperatures \cite{Beian_2015}.

Having characterised the limitations of our approach, we  performed a last experiment under optimised conditions. Precisely, we applied a smooth square  pulse, i.e. an erf-function of 50 ns duration and corrected by the inverse transfer function, $\sim$70 ns after extinction of the loading laser. By comparing the situation where the pulse has 0.4V amplitude to a reference case with zero amplitude, Fig. 4.a shows that the photoluminescence energy reproduces quantitatively the targeted variation, $E_X$ increasing by about 2 meV when the voltage pulse is applied. At the same time, Fig.4.b shows that the photoluminescence intensity does not vary significantly between the two measurements. Thus, we verify that the electrical pulse does induce a sizeable transient current, so that the photoluminescence intensity has no transient variations and is also comparable to its reference value throughout the explored delays to the termination of the loading laser pulse. This highlights that the electrostatic environment is not altered in these experiments, e.g. that the concentration of excess carriers interacting with indirect excitons does not vary significantly.

To summarize, we have shown that electrostatic traps are possibly varied up to GHz frequencies, without inducing large perturbations even at sub-Kelvin bath temperatures. This degree of control is accessible by applying electric pulses directly onto the gate electrodes making an electrostatic trap. Experimentally, we have verified that highest performances require neither impedance matching from an analog waveform generator to the sample heterostructure, nor coaxial cables for the cryogenic environment. Indeed, our approach only relies on the coupling between the dipole moment of indirect excitons and the electric field evanescent inside a field-effect device. It is then sufficient to ensure that the waveform of the signal is faithfully transferred to the surface of the device. The technique that is introduced here then opens the way towards hydrodynamical studies of cold indirect excitons, for instance to explore evaporative cooling schemes and thus quantify the role of exciton-exciton interactions in the thermalisation to sub-Kelvin temperatures.

\begin{figure}[h!]
\centerline{\includegraphics[width=.5\textwidth]{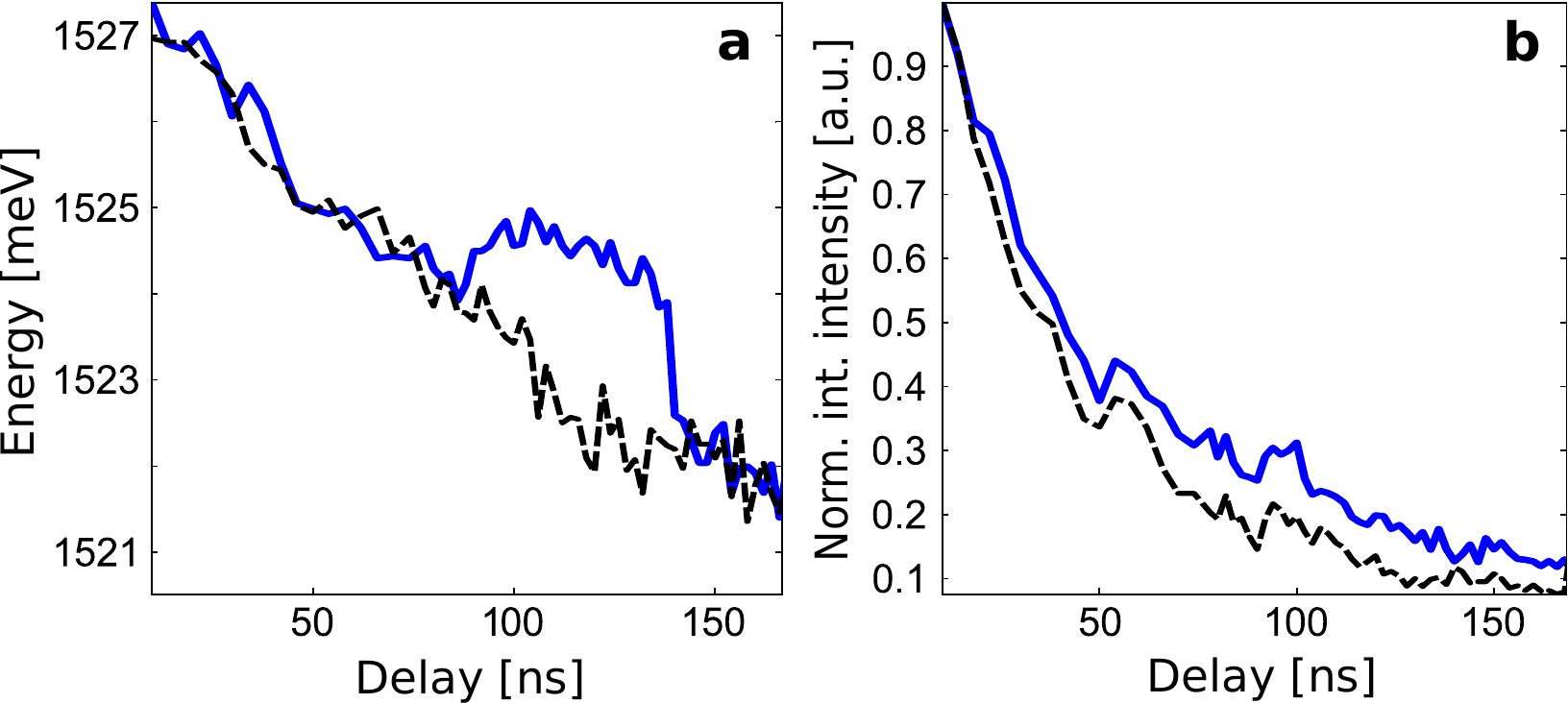}}
\caption{(a) Variation of the photoluminescence energy when a 50 ns long erf-function, with a temporal profile corrected by the inverse transfer function, is applied onto the central trap gate. The solid blue line shows the response measured when the pulse has 0.4V amplitude while the dashed black line shows the same measurements when the amplitude is set to 0. The panel (b) displays the corresponding integrated intensity of the photoluminescence emitted from the trap. Measurements were all realized at a bath temperature set to 330 mK.}
\end{figure} 

\textbf{Acknowledgements:} Our work has been financially supported by the projects XBEC (EU-FP7-CIG) and by OBELIX from the french Agency for Research (ANR-15-CE30-0020), by MINECO (SEVERO OCHOA Grant SEV-2015-0522, FISICATEAMO FIS2016-79508-P),  by the Generalitat de Catalunya (SGR 874 and CERCA program) and by the Fundacio Privada Cellex. We would also like to thank Johannes Guttingen for stimulating discussions and the electronic workshop of ICFO for assistance. Correspondence and requests shall be sent to F.D. (francois$\_$dubin@icloud.com).

\bibliographystyle{apsrev4-1}
\bibliography{Biblio_dynamically_pulsed_revised}

\end{document}